\def\porb{$P_{\rm orb}$}
\def\pspin{$P_{\rm spin}$}
\def\pdot{$\dot{P}$}
\def\rmag{$R_{\rm mag}$}

\documentclass{iaus}
\usepackage{graphicx}

\title[Cataclysmic Variable Stars] 
{Disc--Magnetosphere interactions in Cataclysmic Variable Stars}

\author[C.~Hellier]{Coel Hellier}
\affiliation{Astrophysics Group, Keele University, Staffordshire ST5 5BG, U.K.
\break email: ch@astro.keele.ac.uk}

\pubyear{2007}
\volume{243}  
\pagerange{325--336}
\date{?? and in revised form ??}
\jname{Star--Disk Interaction in Young Stars}
\editors{J.~Bouvier \&\ I.~Appenzeller, eds.}
\begin{document}

\maketitle

\begin{abstract}
I review, from an observational perspective, the interactions
of accretion discs with magnetic fields in cataclysmic
variable stars. I start with systems where the accretion
flows via a stream, and discuss the circumstances in
which the stream forms into an accretion disc, pointing
to stars which are close to this transition.  I then
turn to disc-fed systems and discuss what we know
about how material threads on to field lines, as 
deduced from the pattern of accretion footprints on
the white dwarf.  I discuss whether distortions of the
field lines are related to accretion torques and the
changing spin periods of the white dwarfs.  I also
review the effect on the disc--magnetosphere interaction
of disc-instability outbursts.   Lastly, I 
discuss the temporary, dynamo-driven magnetospheres
thought to occur in dwarf-nova outbursts, and whether
slow-moving waves are excited at the inner edges
of the disc.  
\keywords{accretion, accretion disks, magnetic fields, 
binaries: close, novae, cataclysmic variables, X-rays: stars}
\end{abstract}

\firstsection 
\section{Introduction}
Given that interactions of an accretion disc with a
magnetosphere are widespread in astrophysics, one can
ask why it is of particular interest to study those
in close binary systems such as the cataclysmic variable
stars.  One answer is that these systems often show
a great range of observational clues.  Periodic and
quasi-periodic behaviour, often on timescales that
are easy to study, is the speciality of
cataclysmic variables, making them prime systems
for advancing our understanding of accretion.

If you
want to see how an emission line varies with the
spin-cycle of a magnetic white dwarf in a CV,
repeated 10 times for reliability, you need only
watch for three hours, and the fossil field of the
white dwarf will not have changed.  In contrast,
the same task for a YSO would take weeks, and the
dynamo-driven field might be changing over that time. 
Further, the field of the white dwarf is more likely
to be a simple dipole, and thus easier to model.

In this review I make an observationally led overview
of the disc--field interactions in cataclysmic
variables.  I consider, first, the nature of an
accretion flow in the presence of a magnetic field,
and whether a disc forms, and then turn to how
the disc interacts with the field.   For a
theoretical account of these topics see Li (1999) or
Frank, King \&\ Raine (2002).  The definitive review of 
cataclysmic variable stars is Warner (1995), while for 
a shorter introduction see Hellier (2001).

One big difference from the situation in YSOs is
that in close binaries the accretion originates
in a stream from the secondary star through the inner
Lagrangian point ($L_{1}$).  In cataclysmic variables
with a highly magnetic white dwarf (exceeding $\sim$\,30 MG)
the white dwarf is phase-locked to the orbit.  In such
stars (called AM~Her stars or polars) the
ballistic stream becomes magnetically controlled and
is channelled onto a magnetic pole of the white dwarf.
Since this situation is least similar to YSOs I will
not deal with it here, though for accounts see 
Wickramasinghe \&\ Ferrario (2000) and Schwope \etal\ (2004). 

In a handful of AM~Her stars the white dwarf is
slightly asynchronous with the orbit, despite a high
magnetic field.  Perhaps these stars have been
knocked out of synchronism by a recent nova eruption
(e.g.\ Schmidt \&\ Stockman 1991). The trajectory of the 
stream will now change on the `beat' cycle between the spin and
orbital cycles, as the relative orientation of the
dipole changes.  Typically the accretion stream
flips from one pole to the other and back on the
beat cycle (10--50 days), though at any one time
the star will look like an AM~Her with magnetically
channelled stream accretion.    For accounts of such
stars see Ramsay \etal\ (2000), Ramsay \&\ Cropper (2002), 
Staubert \etal\ (2003) and Schwarz \etal\ (2005).

The main focus of this review will be the `intermediate
polars', which have weaker fields of 1--10 MG.   In
this regime the field is usually not strong enough to
prevent an accretion disc from forming, but is strong
enough to carve out the inner disc and for the
magnetosphere to dominate the observed characteristics.
The hallmark of IPs is X-ray flux from magnetically
channelled accretion onto the white dwarf, heavily
pulsed at the white-dwarf spin period. In some
systems we also see pulsed polarised light, though
in many IPs the polarised light is too diluted by
other parts of the system to be detected.   For an
excellent introduction to these stars see Patterson (1994).

\section{Disc formation: the case of V2400 Oph}
Under what conditions does a disc form around
a magnetic white dwarf?   If the closest approach
of a ballistic stream from the Lagrangian point 
is further out than the magnetic disruption radius,
\rmag\ (at which the magnetic pressure exceeds the 
ram pressure of the material), then the stream
material will accumulate at the circularisation
radius (where the angular momentum of the orbit
matches that of the $L_{1}$ point), and spread
inwards and outwards until the inner edge meets
\rmag.  However, while this 
condition applies in some
wide binaries such as GK~Per, it doesn't hold 
for the field strengths and orbital periods 
typical of most IPs.  Thus it appears that most
IPs form discs even though the trajectory of
a ballistic stream would enter the magnetosphere. 
The case of V2400~Oph gives clues as
to how this might occur.  

V2400~Oph is the most plausible candidate
for an IP which has no disc.  We know the 
white-dwarf spin period from a clear detection
of a 927-s pulsation in polarised light (Buckley
\etal\ 1997), yet there is no X-ray pulse at
this period.  Instead the X-rays are pulsed 
at the 1003-s beat period between the spin
and orbital cycles (Fig.~1), a clear indication
that the accretion proceeds through a stream
which flips between the two magnetic poles
as their relative orientation changes. 
Indeed, an X-ray beat pulse had previously
been proposed as the main diagnostic of 
stream-fed accretion in an IP (Hellier 1991; 
Wynn \&\ King 1992).

\begin{figure}[t]
\includegraphics[width=\textwidth]{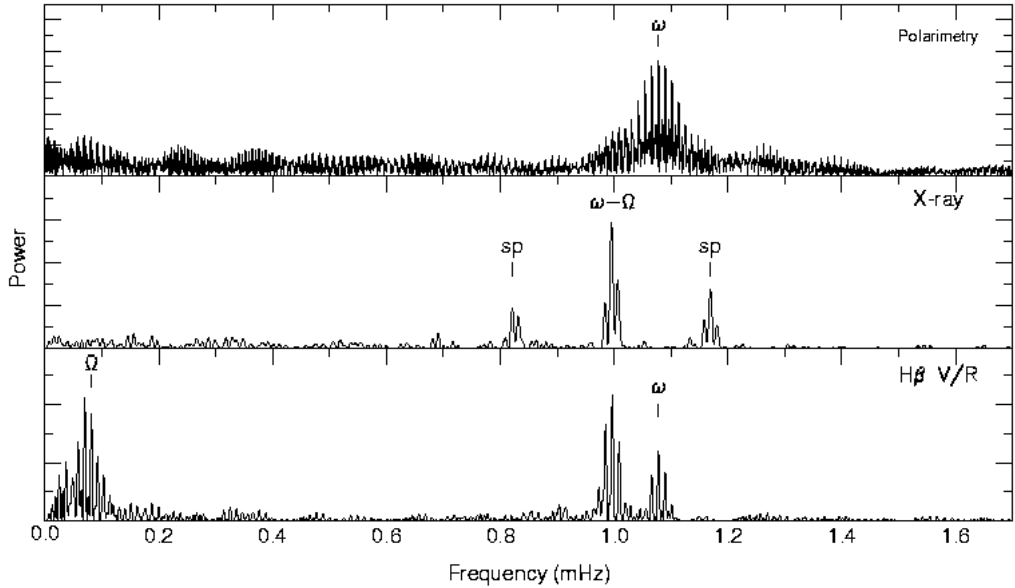}
\caption[]{The Fourier transforms of optical polarimetry,
X-ray flux and emission-line radial velocity for 
V2400~Oph.   The polarimetry reveals the 927-s
spin period (denoted $\omega$) while the X-ray
flux varies instead at the beat period ($\omega$--$\Omega$,
where $\Omega$ is the orbital frequency).   The
emission-line radial velocities vary with all three
periods.  Figure from Hellier (2002), using data
by Hellier and David Buckley.}
\end{figure}

Analysis of how V2400 Oph's spectral lines change
with the spin and beat cycles (Hellier \&\ Beardmore
2002) showed a reasonably good match to a 
simple model of a pole-flipping
stream.  And yet the low amplitudes of the 
pulsations and the absence of any orbital modulation
indicates that this cannot be the whole story. 
Hellier \&\ Beardmore (2002) argued for the presence
of additional orbiting material that is diluting
the emission from the stream.     Yet this 
cannot be an accretion disc, since our experience
of these stars suggests that this would produce an
X-ray spin-period pulsation. 

Instead, we turn to the `diamagnetic blob' scenario
developed by Wynn \& King (1995), which suggests
that dense `blobs' of material can orbit quasi-ballistically,
but with the addition of a `magnetic drag' as they
cross field lines.  Blobs originating in the accretion 
stream could orbit, lose energy as they cross
field lines, spiral inwards and accrete. Yet they
would also tend to screen the field from each other.
The balance of the two factors would determine whether
the blobs accumulate sufficiently to form a disc,
and it appears that V2400~Oph is close to this
borderline, with sufficient orbiting blobs to
carry much of the accretion flow (thus diluting
the stream-fed pulsations), yet with the blobs
not `organised' enough to form a disc or
to produce an X-ray spin pulse. 

V2400~Oph has a field strength (estimated as 9--27 MG;
Buckley \etal\ 1995; V\"ath 1997) that is as high
or higher than in any other IP, and this may explain
why the orbiting blobs cannot quite form a disc in
this star, whereas they do in most IPs.  

A possibly similar case is that of TX~Col.  
Again, the main diagnostic of stream-fed versus
disc-fed accretion is the ratio of the beat-cycle
to the spin-cycle pulsations in the X-ray lightcurve.
In TX~Col the two pulsations are of comparable
magnitude, with the spin pulse being larger at
some epochs and the beat pulse at others (Norton
\etal\ 1997; Wheatley 1999).  So it appears that
in TX~Col the orbiting material has managed to
organise into some sort of disc, yet just as much
of the flow is still in the form of a stream 
flipping between the poles, and the relative 
proportions of the two fluctuate with time. 
Interestingly, Mhlahlo \etal\ 
(2007b) report the detection of high-amplitude 
quasi-periodic oscillations in the optical 
light of TX~Col, at a period of 6000 s that is 
unrelated to the spin or beat periods.  Understanding
these pulsations could be an important clue 
to an accretion flow on the verge of disc
formation. 

\section{EX~Hya-like IPs}
For the magnetosphere to be spinning
in equilibrium with a disc (meaning that the
corotation velocity is close to the Keplerian 
velocity at the inner disc edge) the spin 
period cannot exceed 0.1\,\porb\ (e.g.\ King \& 
Lasota 1991).  This arises since the inner disc edge 
cannot be further out than the circularisation radius,
otherwise the disc would lose more angular
at its inner edge than it gains from the stream,
and it couldn't survive. 

At least three stars, EX~Hya, V1025~Cen and DW~Cnc, 
clearly exceed
this limit, and are instead in a state where the
corotation radius is near the Lagrangian point. 
One possibility is that they are spinning much 
more slowly than equilibrium, perhaps as a result
of having previously been in a discless state
(and it is worth noting that EX~Hya has been 
spinning up monotonically for as long as we've 
been observing it; e.g.\ Hellier \&\ Sproats 1992). 

The alternative possibility is that these stars
currently have no disc.  Although there is not yet
a consensus on this (e.g.\ Hellier \etal\ 1987; 
Hellier, Wynn \&\ Buckley,
2002; Belle \etal\ 2005; Mhlahlo \etal\ 2007a) my
own view is that they probably do have discs.
The main reasons are (1) neither EX~Hya nor
V1025~Cen shows an X-ray
beat pulse in its usual quiescent state, implying
that the accretion loses all memory of orbital 
phase, and (2) neither star shows polarisation,
whereas the discless idea requires them to be 
among the highest-field IPs, with magnetic fields
dominating most of the way to $L_{1}$.  Further,
EX~Hya shows outbursts during which the accretion
stream does overflow the disc and go as far as 
the magnetosphere; the signs of this are obvious,
including an X-ray beat-cycle pulsation, eclipse
profiles of a stream, and high-velocity features in 
the emission lines (e.g.\ Hellier \etal\ 2000), which 
increases our confidence that accretion is purely 
disc-fed in quiescence when these features are not 
seen.\footnote{Mhlahlo \etal\ argue that the 
magnetic influence does go as far as the {\it outer\/} 
edge of the disc in EX~Hya in quiescence.  
The main argument for this is the 
detection of emission with the low velocity
typical of the outer disc which nevertheless 
appears to circle with the spin cycle.    However,
EX~Hya is peculiar in having a spin period 
very close to 2/3rds of the orbit; thus 
orbital-cycle variations do not smear out when
folded on the spin cycle (see Hellier \etal\ 1987).
The feature seen by Mhlahlo \etal\ is simply the usual
{\it orbital\/} S-wave from the edge of the disc,
which contaminates the plots of data folded on the 
spin cycle.}

Why do these three stars with longer-than-expected
spin periods cluster at short periods?   One
explanation might be that in smaller, shorter-period
binaries the synchronisation torques (thought to be
caused by interaction of the magnetic fields of
the primary and secondary) are stronger.  Thus, if a 
low state of no mass transfer leads the disc to
dissipate, a shorter-period system might relatively
quickly head for synchronism whereas a longer-period
system would not.  Once mass-transfer and a disc
are re-established the systems would spin back
up, but they would still have a higher probability
of being found with long, non-equilibrium spin periods.  

\begin{figure}[t]
\includegraphics[width=\textwidth,height=8cm]{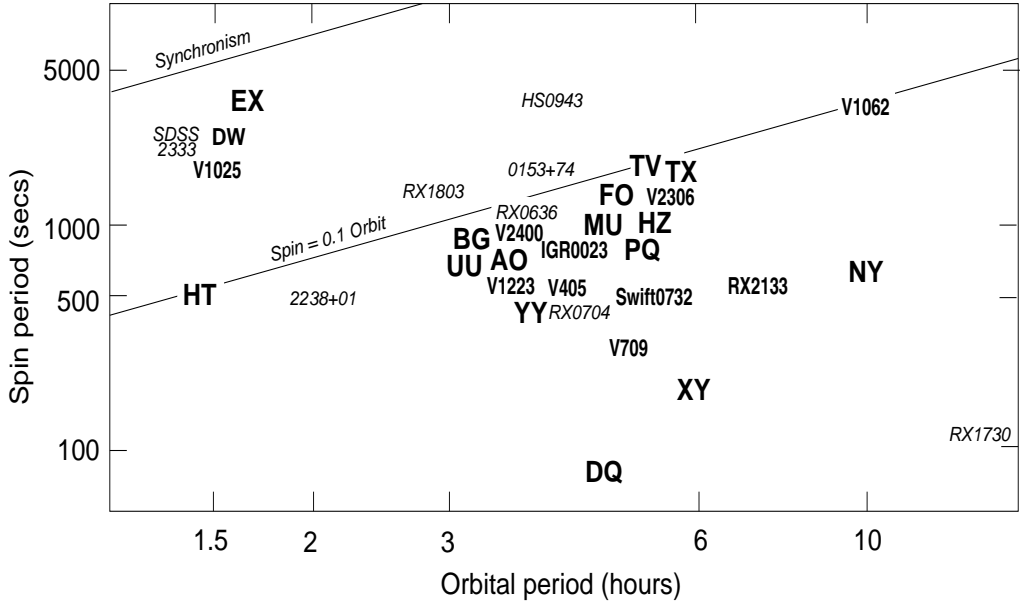}
\caption[]{The spin and orbital periods of the intermediate
polars, indicated by abbreviated names (some labels have 
been moved marginally for clarity).  Italics indicate that 
the location on the plot is uncertain, or that the system is 
not fully secure as an IP, perhaps lacking confirmation from 
multiple datasets or authors.  For details of each system 
see the compilation by Koji Mukai at the `IP home page'  
{\sf http://lheawww.gsfc.nasa.gov/users/mukai/iphome/iphome.html}}
\end{figure}

\section{The disc-fed IPs}
In Fig.~2 I plot the spin and orbital periods of
the known IPs.  I have been conservative in choosing
what to include, placing emphasis on detections
of a periodicity in multiple data sets, preferably
including X-ray data. 

The line at \pspin = 0.1 \porb\ indicates the expected
location for equilibrium rotation of a discless
accretor, such as V2400~Oph.  Systems above this
line cannot both possess discs and be in equilibrium.
Other than the short-period systems just discussed, most
IPs sit on or below this line, compatible with
disc-fed accretion.    Further interpretation of
the \pspin/\porb\ value is
hard without knowledge of the magnetic moment
or the radius of the inner disc edge, neither of
which we have in most IPs.  Norton, Wynn \& 
Somerscales (2004) attempt to extract such information
from the \pspin/\porb\ values, but their method
is predicated on a magnetically dominated, non-disc
accretion flow, and so would not be applicable
to systems with discs.  

Observationally, the evidence is that most
of these systems do have discs, which is deduced
primarily from the fact that the dominant X-ray pulsation
is at the spin period (though see Hellier 1991 for
a discussion of other indicators).  It is also
worth  noting that none of the IPs appear to show
the `soft X-ray excess' that is sometimes seen in
AM~Her stars (see Ramsay \&\ Cropper 2004 for
AM Her stars and Evans \&\ Hellier 2007 for IPs).  
The interpretation of soft excesses in AM Hers
is that blobbiness of the accretion flow can 
survive as far as the white-dwarf surface, and
that such blobs would not undergo a hard-X-ray-emitting
shock but would penetrate the white-dwarf surface
and thermalise to produce soft-X-ray emission (e.g.
Kuijpers \&\ Pringle 1982).  The absence of this
effect in IPs suggests that any such blobs are
destroyed, being shredded in an accretion disc, or 
at least during multiple orbits of the white dwarf.  

\subsection{Stream-overflow in disc-fed stars}
Possession of an accretion disc does
not guarantee that all the accretion flows through
it.   As pointed out by theorists (e.g.\ Lubow 1989) 
the scale height of the stream from the $L_{1}$ point
is likely to exceed that of the outer disc edge,
such that part of the flow continues quasi-ballistically. 
Around the same time, observers were seeing 
direct indications of this (e.g.\ Hellier \etal\ 1989).
In many cataclysmic variables the stream will continue 
inward to some extent,
eventually being subsumed into the disc, but if
the accretion stream flows as far as the 
magnetosphere it will produce a beat-cycle
X-ray pulsation. The relative power in the
beat- and spin-cycle pulsations then presumably
gives an indication of the fractions of material
accreting by the two paths.  

FO~Aqr, AO~Psc, BG~CMi and V1223~Sgr are all
IPs that have a dominant X-ray modulation at 
the spin period, but which show a weaker,
intermittent pulsation at the beat period,
presumably resulting from an overflowing
stream (e.g.\ Hellier 1993; 1998; Norton,
Beardmore \&\ Taylor 1996; Beardmore \etal\ 
1998).  

In addition, observations of the UV lines of 
AO~Psc with {\sl HST\/} have enabled
a more direct observation of the interaction
of the overflowing stream with the field.  
At orbital phases when the stream is in
front of the strong UV backlight of the 
white dwarf, we see narrow absorption dips,
presumably from stream material (Hellier \&\
van Zyl 2005).  These dips show rapid
velocity changes related to the spin
cycle, which reveal the stream flailing
around in accordance with the orientation
of the magnetic field. 

\subsection{Spin-period changes}
In addition to spin periods we can ask about
spin-period changes [see Mukai (2007) for a 
compilation of data].   Long term monitoring
of IPs has found systems where the white dwarf
is spinning up (AO~Psc, BG~CMi, EX~Hya), and systems
where it is spinning down (PQ~Gem, V1223~Sgr), 
and at least one system that shows episodes
of both (FO~Aqr).  Of course we have
monitored these systems for only a small fraction 
of evolutionary timescales, so
it would be dangerous to overinterpret these 
results.  However, the short timescales of
the period changes and the fact that at least one
has swapped from spin up to spin down over a decade
(FO~Aqr; Williams 2003) suggests that a typical IP
below the line in Fig.~2 is hovering about
an equilibrium spin period.  

A decade ago, Patterson (1994) described this issue
as ``murky'' and suggested that a treasure-trove
of \pdot\ data awaited anyone with a good enough
theory to interpret it.  We are still waiting, and 
few authors have been brave enough to go beyond 
Patterson's review. 

\section{The disc--field connection}
One of the hardest questions is what does
the connection between the disc and the
field look like?   What is the radial
extent over which disc material feeds
onto field lines?  What is the vertical
extent?  What does the vertical section
look like?  

We have very few observational
clues, but one approach is to use eclipses, where
the occulting knife edge of the secondary can 
give spatial information on the X-ray-emitting 
accretion regions.  By tracing these
back along field lines, we can deduce 
information about the disc--magnetosphere
interface.  

So far, this technique is only possible in
XY~Ari, which is the only known IP with deep
X-ray eclipses. A study by Hellier (1997b) found 
that the X-ray eclipse egresses occurred in a 
short enough time that the accretion
footprints covered a linear extent of less than 
0.1 white-dwarf radii.  This 
suggests that the feeding onto field lines
is restricted to a relatively small 
azimuthal range.   That is, at least in quiescence;
see Section~6 for the changes wrought by
an outburst.  

Another, more widely applicable technique 
is to deduce the shapes of the accretion footprints 
on the white dwarf by interpreting X-ray spin-pulse 
profiles.

\begin{figure}[t]
\includegraphics[width=0.48\textwidth]{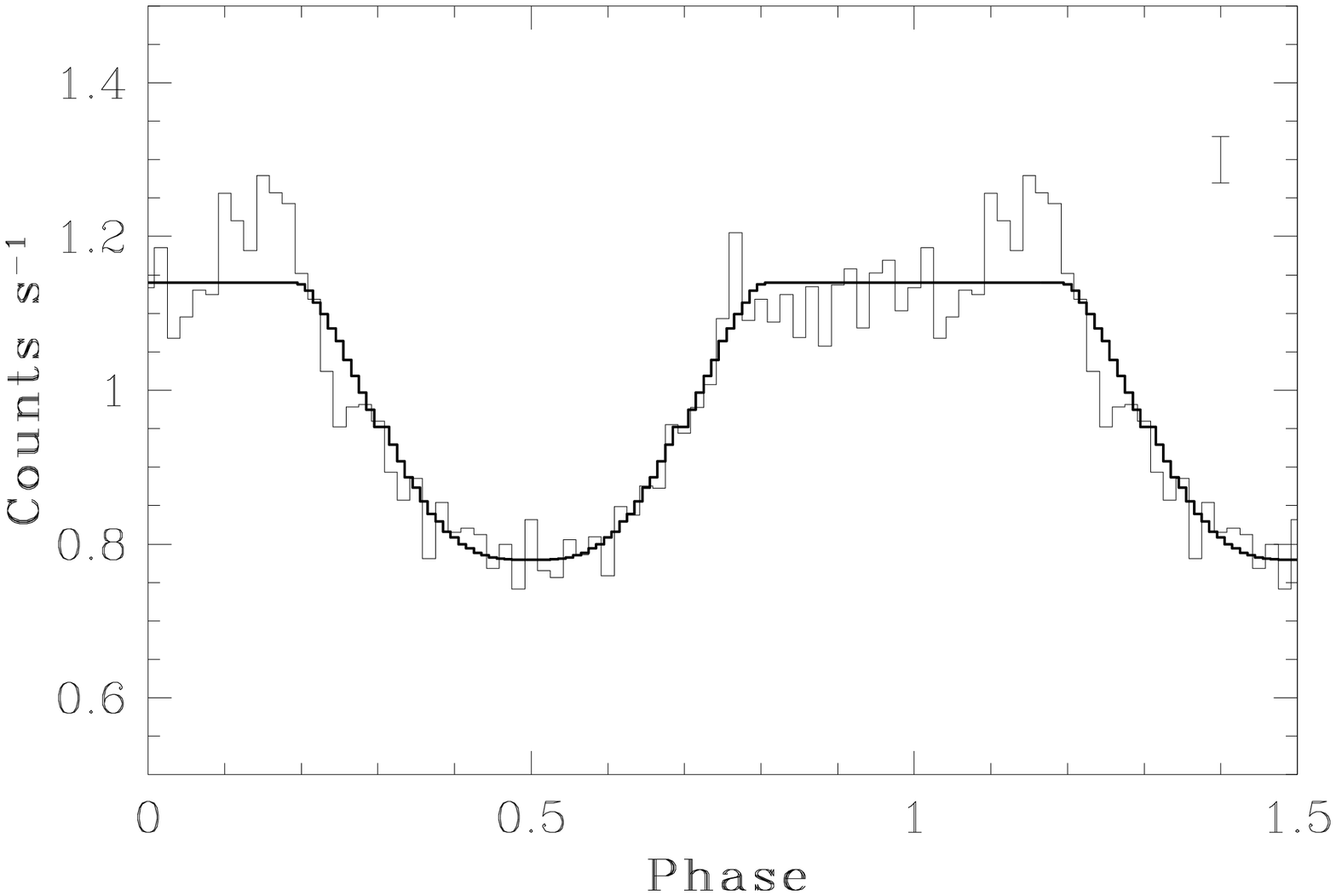}
\hspace{0.04\textwidth}
\includegraphics[width=0.48\textwidth]{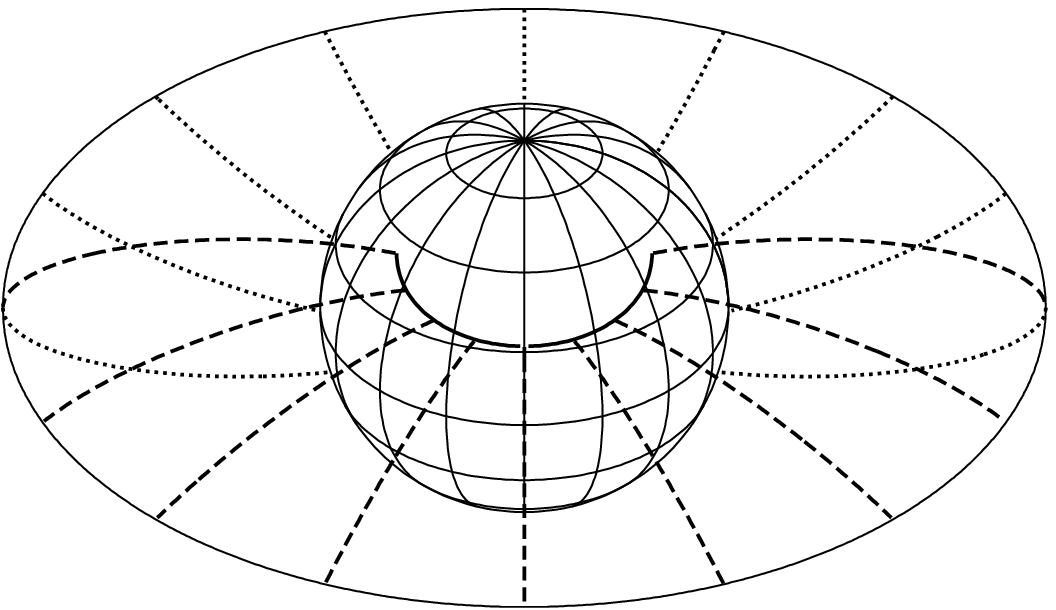}
\caption[]{The `simplest' X-ray spin pulse of HT~Cam.
The left-hand panel shows the X-ray pulse profile and a
fitted model, which is simply a geometric projection
of the accretion regions illustrated in the right-hand
panel.  Figures from Evans \&\ Hellier (2005b). }
\end{figure}

A useful star is HT~Cam, which appears
to have a very simple X-ray spin pulse,
dependent only on whether or not the accretion
footprint is on the visible face of the 
white dwarf.  It shows no phase-varying
absorption, being devoid of the prominent 
absorption dips, produced when the accretion
flow passes in front of the white dwarf,
that are so characteristic of IPs. 

Evans \&\ Hellier (2005b) showed that one
of the simplest possibilities adequately models 
HT~Cam's spin pulse. In the model, which used a
centred dipole, the intensity
of the footprint as a function of magnetic
latitude (which presumably maps to accretion
rate as a function of azimuth at the inner edge
of the disc) was a $\cos^{2}$ function which
peaked at the azimuth to which the magnetic 
pole pointed, falling to zero 90$^\circ$ away,
and then doing the same for the other pole. 

V405~Aur, however, shows a more complex,
double-peaked spin pulse.  In order to explain
this Evans \& Hellier (2004) suggested that the
magnetic dipole was off-centred from the white
dwarf, and that the magnetic axis is highly 
inclined to the spin axis.  

All such work is bedevilled by the number of
parameters that one could vary in a model
(e.g.\ using multipole, non-symmetric fields) 
so it should be cautioned that the proposed
models may not be unique, with more complex
possibilities also possible. 

A similar technique to using X-ray 
pulse profiles is to used spin-pulse profiles
of polarised optical light (e.g. Potter \etal\
1997; 1998).  The idea here is that pulsed
polarised light originates near to the white
dwarf, and so maps to the accretion footprints,
whereas unpolarised optical light, even that
pulsed on the spin cycle, likely originates
from `accretion curtains' much bigger than
the white dwarf.  Where both X-ray and polarimetric 
data combine we can have some confidence in our 
interpretations, even in complex cases such as 
PQ~Gem, discussed next.  

\subsection{Twisting of the field lines}
The `simplest' IP spin pulse of HT~Cam is compatible
with field lines that are undistorted.   However,
detailed analysis of the complex spin
pulse of PQ~Gem, using X-ray data, optical
spectroscopy, and polarimetry, suggests that the
accreting field lines are distorted by accretion 
torques, and that the accreting field lines are 
0.1 cycles ahead of the magnetic
pole (Mason 1997; Potter \etal\ 1997; Hellier 1997a;
Evans, Hellier \&\ Ramsay 2006).  

Similar analysis
of FO~Aqr yields the opposite: the accreting field
lines appear to lag the magnetic pole by a quarter
of a cycle (Evans \etal\ 2004).

An obvious question is whether these twists are
related to the torques on the white dwarf, as
shown by period changes.  Indeed, PQ~Gem, with
field lines swept ahead of the pole, is found
to be spinning down (Mason 1997).  FO~Aqr, with
field lines swept behind the pole, is spinning
up --- at least at some epochs.  But it has also shown
a period of spin down (Williams 2003); and thorough
investigation of the X-ray spin pulse over the
different epochs (e.g.\ Beardmore \etal\ 1998) show
no obvious shifts in the phases of absorption dips
that would imply that the field lines had swapped
from trailing to leading.

Thus there is no simple interpretation of the
current information on field-line twists and
period changes; as yet we have such information
on too few systems to discern patterns.

\section{Disc--Magnetosphere interactions in outburst}
We presume that in many of the IPs the white-dwarf magnetospheres
are rotating close to equilibrium with their discs.   However,
cataclysmic variables also show dwarf-nova outbursts,
where a hydrogen-ionisation instability in the disc
causes a hundred-fold increase in the accretion rate,
lasting typically for several days (e.g.\ Lasota 2001;
Osaki 2005).    In a few cases
we have observed outbursts in IPs, which allows us to
watch the dynamic behaviour of a magnetosphere.

Theory tells us (e.g.\ Frank \etal\ 2002)
that as the ram pressure of the accretion flow
increases the magnetosphere shrinks (as $r \propto
\dot{M}^{-2/7}$).   In an outburst of XY~Ari we saw
confirmation of this, with major changes in the
X-ray pulse profiles indicating that the disc had
pushed inwards and blocked the view of the lower
magnetic pole (Hellier, Mukai \&\ Beardmore 1997).
It took about a day for the disc to push inwards
from $\sim$\,9 $R_{\rm wd}$ to $\sim$\,4 $R_{\rm wd}$,
and eclipse timings indicate that, when it had done so,
the flow of material onto field lines was no longer
restricted in azimuthal extent, but now flowed from
all parts of the inner disc, to fill a complete
ring of magnetic longitude around the magnetic poles.
It is unclear whether this change is due more to the
non-equilibrium rotation (presumably after the disc
has pushed inward the Keplerian velocity at its
inner edge would now exceed the speed at which the
magnetosphere rotates) or whether it relates to the
greater scale height of the disc during outburst.

GK~Per has an exceptionally long orbital period,
and thus a very large disc with long-lived outbursts
that are thus easy to study.  Analysis of the X-ray
pulse profiles suggests that GK~Per behaves
similarly to XY~Ari.  The pulse-profiles are
explainable by the disc pushing inwards and
cutting off the view of the lower pole.  Again,
the accretion appears to feed from all disc azimuths
at the height of outburst, causing complete rings
of accretion at the magnetic poles (Hellier, Harmer
\&\ Beardmore 2004; see also Vrielmann, Ness \&\
Schmidt 2005).

Thus it is clear that observations of IPs in
outburst can reveal important observations of the
dynamic interaction of a disc and magnetosphere.
However, the fact that such outbursts are occasional,
short-lived and unpredictable makes such data
hard to obtain.  In addition to XY~Ari and GK~Per,
YY~Dra (Szkody \etal\ 2002) and HT~Cam (Ishioka \etal\
2002) are
among the few systems where this has been achieved.
Some IPs (EX~Hya, V1223~Sgr \&\ TV~Col) show even
shorter-lived outbursts, lasting only a day, that
are even harder to study.  So far it is unclear whether
these are disc-instability outbursts (shorter-lived
and lower-amplitude owing to the magnetic truncation
of the disc) or whether they are episodes of mass
transfer from the secondary star (see, e.g., Hellier \etal\
2000 and references therein).

\section{Temporary magnetospheres: DNOs}
A long-studied feature of dwarf-nova outbursts
is the occurrence of `dwarf-nova oscillations'
in the lightcurves. These are semi-stable periodicities
with periods of 6 to 50 secs, which are presumably
related to the white-dwarf spin.   The puzzle has been
why, if they indicate a spinning magnetosphere, are
they seen only in outburst in what are otherwise
considered to be non-magnetic systems?  Surely the magnetic
field should be most manifest in quiescence, when
the accretion rate is least?  The resolution proposed
in the series of papers Warner \& Woudt (2002), Warner \&\ Woudt 
(2002), Warner, Woudt \&\ Pretorius (2003) is that
the field is only present in outburst, being generated
by a dynamo caused by an equatorial belt which is spun up 
by the extra accretion in outburst and is slipping over the
body of the white dwarf.  The low-coherence of the
oscillations is then explained by the low inertia of
the equatorial belt.

The phenomenology of DNOs is too vast to summarise
here (see Warner 2004, Pretorius, Warner \& Woudt 2006 
and references therein), but an important topic is their
link to the `quasi-periodic oscillations' sometimes
seen in dwarf novae.

\section{QPO waves in the inner disc?}
Quasi-periodic oscillations are longer-period than
DNOs and less coherent (for an account of the
phenomenology see Warner 2004).  Although they have been
known for decades, it is only relatively recently
that a plausible explanation has been proposed.
Warner \&\ Woudt (2002b) suggest that QPOs are caused
by travelling waves at the inner edge of the disc,
excited by the interplay of the dynamo-generated
field with the inner disc.  The waves are slow-moving,
prograde bulges that modulate the observed light by
periodically obscuring the white dwarf as they circle.
Empirically, they appear to circle $\approx$\,15 times
more slowly than the DNO period, which is presumably
the rotation period of the temporary, dynamo-driven magnetosphere.

Such bulges can also produce a phenomenon
by re-processing the light from the DNO, leading to
a quasi-periodicity at the beat period of the DNO
and the QPO (see Woudt \&\ Warner 2002; Marsh \&\
Horne 1998).

\begin{figure}[t]
\hspace*{0.1\textwidth}\includegraphics[width=0.8\textwidth]{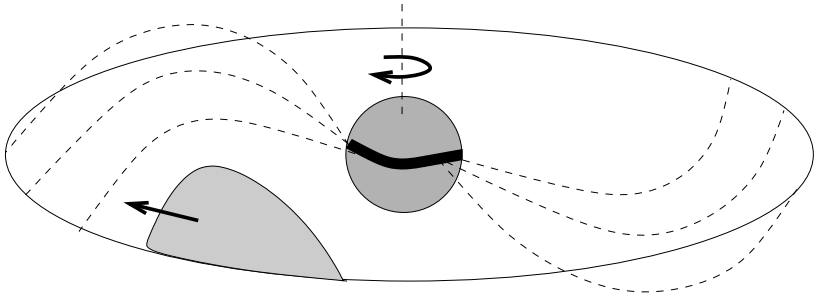}
\caption[]{A schematic illustration of a dwarf nova
in outburst.  While these systems are non-magnetic
is quiescence, the increased accretion of outburst
spins up an equatorial belt.  This slips over the
white dwarf, forming a dynamo and generating a 
temporary magnetosphere.  At the inner edge of the
disc slow-moving bulges are produced by prograde 
travelling waves.}
\end{figure}

An interesting question is whether these travelling
waves are peculiar to the conditions of a dwarf-nova
outburst, or whether they are widespread in disc--magnetosphere
interactions.   Warner \etal\ (2003) present empirical
evidence that the $P_{\rm QPO} = 15 P_{\rm DNO}$
relation holds in a wide range of objects, from neutron
stars to black holes.  We can thus ask whether the
phenomenon is also present in IPs, with their
permanent, fossil fields.

One IP almost certainly showing this phenomenon
is GK~Per, which in outburst shows large-amplitude
absorption dips, recurring quasi-periodically with
a 5000-s timescale, 15 times longer than the 351-s
spin period (Watson, King \&\ Osborne 1985).

One proposed model (e.g.\  Morales-Rueda, Still \&\
Roche 1999) suggests that the 5000-s timescale is
the beat period between the 351-s spin period and
the Keplerian velocity at the inner disc edge.
However Hellier \&\ Livio (1994) show that,
phenomenologically, the modulation must be caused
by periodic obscuration by structure circling
at 5000 s.  Such structure is then best explained
by the Warner \&\ Woudt (2002) idea.

Are the travelling-wave bulges there only in
the non-equilibrium conditions of outburst,
or are they present also in quiescence?   It is
hard to tell.  As discussed in Hellier, Harmer
\&\ Beardmore (2004), the inner disc edge
will be $\sim$\,5 times further out in quiescence,
meaning that, at the inclination of GK~Per, the 
bulges would not obscure the white dwarf and
create absorption dips.

Thus, it may be that we see the QPOs readily
enough in dwarf novae, where the weak, dynamo-driven
magnetospheres place the bulges near to the white
dwarf, where obscuration effects will be easily
seen.  But in IPs, with stronger fields and
larger magnetospheres, the inner disc edges
are too far from the white dwarf for obscuration
by bulges to be obvious.

It is also worth noting that in most IPs
a quasi-periodicity at 15 times the spin
period would be close to the orbital cycle,
and thus hard to distinguish from orbital-cycle
modulations, which are often prominent (e.g.\
Hellier, Garlick \&\ Mason 1993; Parker, Norton 
\&\ Mukai 2005).

Also, given that QPO effects are usually subtle,
they would require trains of data of 10 cycles
or more to detect, and this is a timescale
(30 hrs) on which it is very hard to obtain
continuous data trains.  Thus for practical
reasons, the question of how widespread this
phenomenon is in IPs is very hard to answer.

\section{Concluding remarks}
I have shown above that in many cases observations
of magnetic cataclysmic variables give robust
clues to how an accretion disc interacts
with a magnetosphere.  For completeness, I briefly
mention two further scenarios.  First, there is 
the possibility of a magnetosphere spinning 
sufficiently fast to expel material, forming
a `propeller' system.  AE~Aqr is thought to be
in this state, with expulsion of material powered
by a rapid spin-down of the white dwarf (e.g.\
Wynn, King \& Horne 1997; Meintjes \& de Jager 2000;
Meintjes \&\ Venter 2005).

Another possibility is that of a white dwarf with
a misaligned spin axis, which is precessing.   This
would introduce a further periodicity
at the precession period into the panoply of 
cataclysmic-variable periodicities.  This has been 
invoked for stars such as FS~Aur and HS\,2331+3905
that show periodicities which are otherwise very
hard to explain (e.g.\ 	Tovmassian \etal\ 2003; 
Araujo-Betancor \etal\ 2005; Tovmassian, Zharikov \&\ 
Neustroev 2007).  

One notable feature, however, is that owing to
the diversity of phenomena even within the intermediate
polars, we are often interpreting observational features
seen in only one or two systems, rather than 
analysing patterns seen throughout the class.  

There is only one propeller system (AE~Aqr); 
FS~Aur and HS\,2331+3905 are both unique; only
one system (V2400~Oph) is clearly discless; 
only two systems (EX~Hya and GK~Per) are well
studied in outburst, and both are very different
from each other. We have only one system with deep
X-ray eclipses (XY~Ari, hidden behind a molecular cloud
where it cannot be studied in the optical) and one
grazing eclipser (EX~Hya).    Studies of accretion-curtain
twists and their relation to spin-period changes are 
available for only a couple of systems (FO~Aqr and
PQ~Gem); studies of the formation mechanism of the
X-ray spin pulses are available for half a dozen 
stars, but they show a wide range of behaviours,
some being single pulsed at the white-dwarf spin
period and others being double pulsed (e.g.\ Evans
\&\ Hellier 2005a).  

Thus, progress in understanding disc--magnetosphere 
interactions in these stars is likely to come from 
studying sufficient systems in detail to
look for patterns encompassing the class. Assisting
this is the fact that, in the past 5 years, fully a 
dozen objects have been added to the number we can
plot on Fig.~2.  Most of these are only lightly studied,
and thus the prospects for learning much more about
disc--magnetosphere interactions from this class of
objects are bright.

\end{document}